\documentclass[prd,preprint,superscriptaddress,showpacs,byrevtex]{revtex4}
\usepackage{epsfig}

\newcommand{\be}{\begin{equation}}
\newcommand{\ee}{\end{equation}}
\newcommand{\bea}{\begin{eqnarray}}
\newcommand{\eea}{\end{eqnarray}}
\def\simgt{\rlap{\lower 3.5 pt \hbox{$\mathchar \sim$}} \raise 1pt
Ê \hbox {$>$}}

\begin{document}

\title{Child universes UV regularization?}

\author{E.I. Guendelman } \email{guendel@bgu.ac.il}
 \affiliation{Physics Department, Ben-Gurion University of the Negev,
Beer-Sheva 84105, Israel}

\date{\today}

\begin{abstract}
It is argued that high energy density excitations, responsible for
UV divergences in quantum field theories, including quantum
gravity, are likely to be the source of child universes which
carry them out of the original space time. This decoupling
prevents these high UV excitations from having any influence on
physical amplitudes. Child universe production could therefore be
responsible for UV regularization in quantum field theories which
take into account gravitational effects. Finally child universe
production in the last stages of black hole evaporation, the
prediction of absence of tranplanckian primordial perturbations,
connection to the minimum length hypothesis and in particular the
connection to the maximal curvature hypothesis are discussed.

\end{abstract}
\maketitle

\newpage

\section{Introduction}

Quantum field theory and quantum gravity in particular suffer from
UV divergences. While some quantum field theories are of the
renormalizable type, quantum gravity is not and the UV divergences
cannot be hidden into a finite number of "counter-terms".
Perturbative renormalizability does not appear to be available for
quantum gravity.

In an apparently unrelated development, the "child universe"
solutions have been studied  \cite{blau}, \cite{ansoldi}. These
child universes are regions of space that evolve in such a way
that they disconnect from the ambient space time. Inflationary
bubbles of false vacuum correspond to this definition \cite{blau},
\cite{ansoldi}. In this case an exponentially expanding
inflationary bubble arises from an ambient space time with zero
pressure which the false vacuum cannot displace. The inflationary
bubbles disconnect from the ambient space generating a child
universe.

Here we want to explore the possibility that high energy density
excitations, associated to the UV dangerous sector of quantum
field theory could be the source of child universes, which will
carry the UV excitations out of the original ambient space time.
Child universe production could be therefore responsible for UV
softening in quantum field theory that takes into account
gravitational effects. It implies also the existence of a maximum
energy density and curvature.

We will now show now, using a simple model, that very high UV
excitations have appreciable tendency to disconnect from the
ambient space time

\section{The Super high UV Bubble}

We  describe now the model which we will use to describe a high UV
excitation which will be associated with  the production of a
child universe. This model for high UV excitation will consist of
a bubble with very high surface tension and very high value of
bulk energy density inside the bubble.

The entire space-time region consists of two regions and a
boundary: 1) {\bf Region I} de Sitter space 2) {\bf Region II},
Schwarzschild space and  the domain wall boundary separating
regions I and II.

In {\bf Region I}: The de Sitter space. The line element is given
by

\bea ds^2 =-(1-\chi^2 r^2) dt^2 +(1-\chi^2 r^2)^{-1} dr^2 + r^2
d\Omega^2 \label{de} \eea where $\chi$ is the Hubble constant
which is given by \bea \chi^2 = \frac{8}{3} \pi G \rho_0
\label{chi4} \eea

$\rho_0$ being the vacuum energy density of the child universe and
$ G=\frac{1}{m_P^2} $ where $m_P=10^{19}$ GeV.

In {\bf Region II}: The Schwarzschild line element is given by
\bea ds^2 = -(1-\frac{2GM}{r})dt^2 +(1-\frac{2GM}{r})^{-1}dr^2
+r^2 d\Omega^2 \label{ma} \eea

The Einstein`s field equations,  \bea R_{\mu \nu} -\frac{1}{2}
g_{\mu \nu} R =  8 \pi G T_{\mu \nu}. \label{ein} \eea are
satisfied in regions I and II and determine also the domain wall
evolution \cite{blau}, using the methods developed by Israel
\cite{israel}. Using gaussian normal coordinates, which assigns to
any point in space three coordinates
 on the bubble and considers then a geodesic normal to the bubble which
 reaches any given point after a distance $\eta$
 ( the sign of $\eta$ depends on  which side of the bubble the point is
 found). Then energy momentum tensor $T_{\mu \nu}$ is given by \bea
T_{\mu \nu}(x) && = -\rho_0 g_{\mu \nu}, ~~~~~~~~~~~~~~~~(\eta < 0){\rm ~for~ the~ child~ universe, ~~~~{\bf Region ~I}, ~(negative~pressure)} \nonumber \\
&& =0, ~~~~~~~~~~~~~~~~~~~~~~~~(\eta > 0){\rm for~ the~ Schwarzschild ~~region, ~~~~{\bf Region II}} \nonumber \\
&&= - \sigma h_{\mu \nu} \delta(\eta)
~~~~~~~~~~~~~~~~~~~~~~~~~~~~~{\rm for ~ the
~domain~wall~boundary}. \label{tmunu} \eea

where $\sigma$ is the surface tension and $h_{\mu \nu}$ is the
metric tensor of the wall, that is $h_{\mu \nu} = g_{\mu
\nu}-n_{\mu}n_{\nu} $,  $n_{\mu}$ being the normal to the wall.

The eq. of motion of the wall give\cite{blau} \bea M_{\rm cr} =
\frac{1}{2G \chi} \frac{\gamma^3 z_m^6
(1-\frac{1}{4}\gamma^2)^{\frac{1}{2}}}{3 \sqrt{3} (z_m^6
-1)^{\frac{3}{2}}} \label{cr} \eea where the $M_{\rm cr}$ is the
mass at (or above) which there is classically a bubble that
expands to infinity into a disconnected space, the child universe.
In the above equation \bea
&& \gamma = \frac{8 \pi G \sigma }{\sqrt{\chi^2 + 16 \pi^2 G^2 \sigma^2}} \nonumber \\
&& z_m^3 = \frac{1}{2} \sqrt{8+(1-\frac{1}{2}\gamma^2)^2} -
\frac{1}{2} (1-\frac{1}{2}\gamma^2) \label{gm} \eea

where $z^3 = \frac{\chi_{+}^2 r ^3}{2GM}$ and $\chi_{+}^2 = \chi^2
+\kappa^2 $, $\kappa = 4 \pi G \sigma$. $r_m$ is the location of
the maximum of the potential barrier that prevents bubbles with
mass less than $M_{\rm cr}$ to turn into child universes.

We expect this representation of a high UV excitation to be
relevant even for a purely gravitational excitation, which can be
associated , after an appropriate averaging procedure, to an
effective energy momentum, a procedure that gets more and more
accurate in the UV limit.

Let us now focus our attention on the limit where $\sigma
\rightarrow \infty$  (while $\rho_0$ is fixed), which we use as
our first model of a super UV excitation. Then, we see that
$\gamma \rightarrow 2$ and $M_{\rm cr} \rightarrow  0$.
Alternatively, we could obtain another model of super UV
excitation, by considering the energy density inside the bubble,
$\rho_0 \rightarrow \infty$ , while keeping $\sigma$ fixed. This
also leads to $M_{\rm cr} \rightarrow 0$ as well. Finally, letting
both $\sigma \rightarrow \infty$ and $\rho_0 \rightarrow \infty$
while keeping their ratio fixed, also leads to $M_{\rm cr}
\rightarrow  0$. In all these limits we also get the the radius of
the critical bubble $r_m \rightarrow 0$ as well.

In \cite{blau} the above expression for $M_{\rm cr}$ was explored
for the case that energy densities scales (bulk and surface) were
much smaller than the Planck scale, like the GUT scale. This gave
a value for $M_{\rm cr} = 56 kg >>m_p $. Here we take the
alternative view that the scale of the excitations are much higher
than the Planck scale, giving now an arbitrarily small critical
mass. Defining the "scale of the excitation" through by $\rho_0
\equiv M_{exc}^4 $, then the pre-factor $\frac{1}{2G \chi}$ in eq
(\ref{cr}), goes like $\left( \frac{m_p}{M_{exc}} \right)^2 m_p $.
We see that for trans planckian excitations, i.e. if
$M_{exc}>>m_p$, we obtain a very big reduction for $M_{\rm cr}$.
This is a kind of "see saw mechanism", since the higher the
$M_{exc}$, the smaller $M_{\rm cr}$.

This means that in these models for high UV excitations there is
no barrier for the high UV excitation to be carried out to a
disconnected space by the creation of a child universe. Notice
also the interesting "UV -IR mixing" that takes place here:
although we go to very high UV limits in the sense that the energy
density in the bulk or the surface energy density are very high,
the overall critical mass goes to zero.

\section{The Conjecture}
This allows us to formulate the conjecture that the dangerous UV
excitations that are the source of the infinities and the non
renormalizability of quantum gravity are taken out of the original
space by child universe production, that is, the consideration of
child universe production in the ultrahigh (trans planckian)
sector of the theory could result in a finite quantum gravity,
since the super high UV modes, after separating from the original
space will not be able to contribute anymore to physical
processes.

The hope is that in this way, child universes could  be a of
interest not only in cosmology but could become also an essential
element necessary for the consistency of quantum gravity.  One
situation where all the elements required (high energy densities ,
since the temperature is very big) necessary for obtaining a child
universe appears to be the late stages of Black Hole evaporation.
If the ideas explained here are correct, we should not get
contributions to primordial density perturbations from the trans
planckian sector, since these perturbations would have
disconnected from our space time. Also, any attempt to measure
distances smaller than the Planck length will be according to this
also impossible since such a measurement will involve exciting a
high UV excitation that will disconnect. This means that there
must be a minimum length that we could measure, of the order of
the Planck scale.

It appears there is a maximal energy density according to this,
since now bubbles with high energy density will be quickly
disconnected, being replaced  in the observable universe by
regions of Schwarzschild space, which has zero energy density,
i.e., a very big energy density must decay in the observable
universe. The "maximal curvature"\cite{maxcurv} hypothesis (here
we focus on scalar curvature) is justified by this maximal energy
density result, if we use eq. (\ref{ein}). An effective dynamics
that takes into account the effect of child universe production
(i.e. integrates out this effect) could resemble indeed that of
\cite{maxcurv}. Notice that the maximal scalar curvature
hypothesis gives rise to very interesting dynamics\cite{easson}.

 \acknowledgements
I would like to thank G. Nayak for help in the composition of this
paper and for very useful discussions. I also want to thank S.
Ansoldi for interesting communications and to D. Easson for
discussions related to refs. \cite{maxcurv},\cite{easson}.

\end{document}